# Entropy stabilization and effect of *A*-site ionic size in bilayer nickelates

Jia-Yi Lu[1], Jia-Xin Li[1], Xin-Yu Zhao[1], Ya-Nan Zhang[1,2], Yi-Qiang Lin[1], Kai-Xin Ye[1,2], Hui-Qiu Yuan[1,2,3,4,] and Guang-Han Cao[1,3,4*]

The discovery of high-temperature superconductivity in bulk samples of $La_3Ni_2O_{7-\delta}$ (La-327) nickelate under high pressure [1-5] and in La-327 thin films at ambient pressure [6-8] opened up a new chapter in superconductivity research. This material is an $n = 2$ member of the Ruddlesden-Popper (RP) series $A_{n+1}Ni_nO_{3n+1}$, where $n$ is the number of perovskite $ANiO_3$ layers in between rocksalt-type $A$O layers. Previous studies have shown that La-327 has a relatively narrow phase stability range [9-11], and the fact that merely one member of the bilayer nickelate (i.e. La-327) has been synthesized so far suggests its inherent instability. Such inherent instability often leads to stacking faults which tend to destroy bulk superconductivity under high pressure [12].

To address these challenges, considerable research efforts have been devoted to the phase stabilization via chemical substitutions [13-18]. We found that aluminum doping effectively enhances the stability of La-327, although it destroys both pressurized superconductivity and the density-wave (DW) order [13]. Doping at the $A$ site with rare-earth elements, such as Pr, has been shown to significantly improve the phase purity and reduce the stacking faults in the 327-type compounds [7,8,12]. Subsequent studies [14-18] have further demonstrated that reducing the average $A$-site ionic radius, $\bar{r}_A$, leads to lattice contraction and enhances the orthorhombic distortion, an effect distinct from that of physical pressure [15,19]. These findings suggest that incorporating smaller $A$-site cations not only shrinks the lattice effectively but also enhance the pressurized superconductivity, in agreement with theoretical predictions [19,20].

The high-entropy (HE) strategy offers a promising way to stabilize a given phase. This strategy involves incorporating multiple elements into a specific crystallographic site, thereby effectively stabilizes the structure. The concept of HE materials originated from earlier works on HE alloys [21,22], and has led to widespread research interest due to their exceptional functional properties and stability [23-26]. HE oxides exhibit tunable electronic structures and distinctive physical properties [27]. A key metric for classifying oxides is configurational entropy, $S_{\mathrm{conf}} = -R\sum_s m_s \sum_i x_i \ln x_{i,s}$, where $R$ is the gas constant, $m_s$ is the multiplicity of sublattice $s$, and $x_{i,s}$ is the fraction of element $i$ on sublattice $s$ [27,28]. According to this parameter, oxides are grouped into low-entropy oxides ($S_{\mathrm{conf}} \leq 1.0R$), medium-entropy (ME) oxides ($1.0R < S_{\mathrm{conf}} < 1.5R$), and HE oxides ($S_{\mathrm{conf}} \geq 1.5R$) [27]. To date, numerous HE oxides with various crystal structures have been successfully synthesized [29]. Despite these achievements, entropy-stabilized bilayer RP-type nickelates remain scarcely explored. Thus, the application of the HE approach is expected to expand the compositional space of the bilayer nickelate family further, potentially optimizing the pressurized superconductivity.

In this work, we successfully synthesized medium-entropy $La_{1.2}Pr_{0.6}Nd_{0.6}Sm_{0.6}Ni_2O_{7-\delta}$ (ME-327) and high-entropy $La_{0.67}Pr_{0.67}Nd_{0.67}Sm_{0.33}Eu_{0.33}Gd_{0.33}Ni_2O_{7-\delta}$ (HE-327). The experimental methods are provided in Supplementary Information (SI). The occupations in the $A$1 and $A$2 sites [see inset in Fig. 1(a)] were designed in order to stabilize the 327 phase with small average $A$-site ionic radius, $\bar{r}_A = \sum_i x_i r_i$, where $r_i$ is the ionic radii for 9-fold coordination and $x_i$ is molar fraction satisfying $\sum_i x_i = 1$, together with maximization of the configuration entropy. The $\bar{r}_A$ values of ME- and HE-327 are 1.181 and 1.164 Å, close to those of hypothetical $Pr_3Ni_2O_7$ and $Nd_3Ni_2O_7$, respectively. Meanwhile, these compositions satisfy the ME and HE criteria, regardless of random or partially preferential occupations (Table S1 in the SI) [30,31].

Polycrystalline samples of La-, ME-, and HE-327 were all single phase, as shown in Fig. 1(a) and Fig. S1. The successful synthesis of 327 phases with $\bar{r}_A$ values close to those of $Pr^{3+}$ and $Nd^{3+}$ indicates entropy stabilization in bilayer nickelates. The widths of the XRD reflections of ME and HE samples are as sharp as those of La-327, indicating highly good crystallinity. The Rietveld refinements using the *Amam* model converged well, with reliable factors of $R_{\mathrm{wp}} = 5.20\%$ and the goodness of fit of $S = 1.68$ for HE-327. The detailed crystallographic data are presented in Table S2. Fig. 1(b) shows a systematic shift of the XRD peaks with decreasing $\bar{r}_A$. The (006) and (200) reflections shift to higher $2\theta$ angles, reflecting contractions along $a$ and $c$ axes, while the (020) peak shifts to lower angles, indicating an expansion along the $b$ axis. This means an enhanced orthorhombic distortion. The lattice parameters are plotted in Fig. 1(c). Compared with 'unalloyed' La-327 [Fig. S1 (a)], the *a* axis decreases by 0.9% and 1.3%, and the *c* axis is 1.6% and 2.4% reduced, for ME- and HE-327, respectively. Notably, the lattice contraction in our HE-327 sample exceeds that of $La_{0.6}Nd_{2.4}Ni_2O_{7-\delta}$ [18], thereby exerting greater chemical pressure in the bilayer nickelate system.

[1] School of Physics, Zhejiang University, Hangzhou 310058, China
[2] New Cornerstone Science Laboratory, Center for Correlated Matter and School of Physics, Zhejiang University, Hangzhou, China.
[3] Institute of Fundamental and Transdisciplinary Research, and State Key Laboratory of Silicon and Advanced Semiconductor Materials, Zhejiang University, Hangzhou, China.
[4] Collaborative Innovation Centre of Advanced Microstructures, Nanjing University, Nanjing, 210093, China
[*] Corresponding author (email: ghcao@zju.edu.cn)

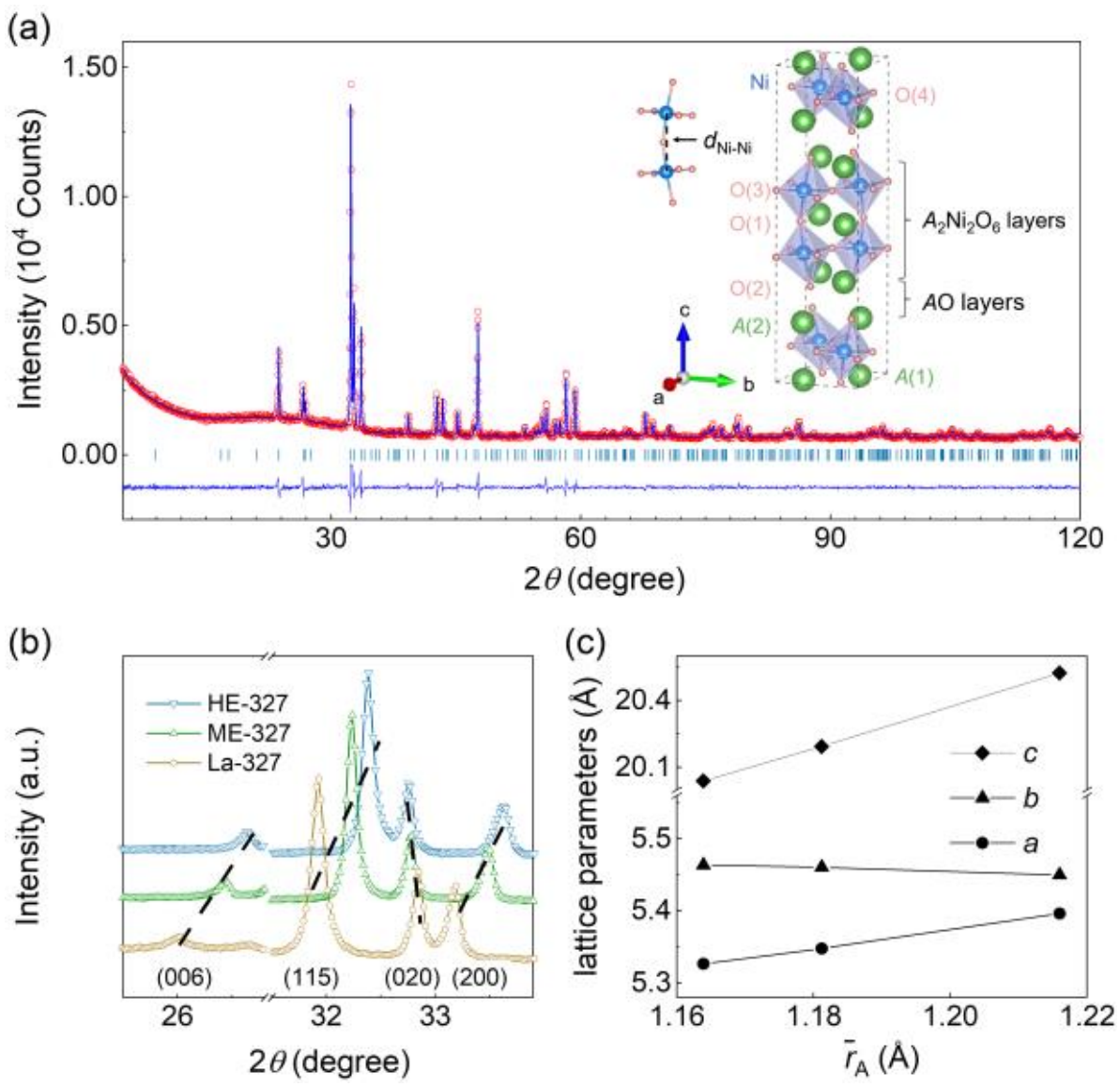

**Figure 1.** (a) Rietveld refinements of the powder XRD diffractions of HE-327 sample. The inset shows the crystal structure of $A_3Ni_2O_7$ (*A* is a mixture of trivalent rare-earth elements), consisting of alternating perovskite $A_2Ni_2O_6$ layers and rocksalt-type *A*O layers. (b) Closeups of the XRD patterns of La-, ME-, and HE-327, showing shifts of the representative reflections. (c) Lattice parameters as functions of the average ionic radius, $\bar{r}_A$.

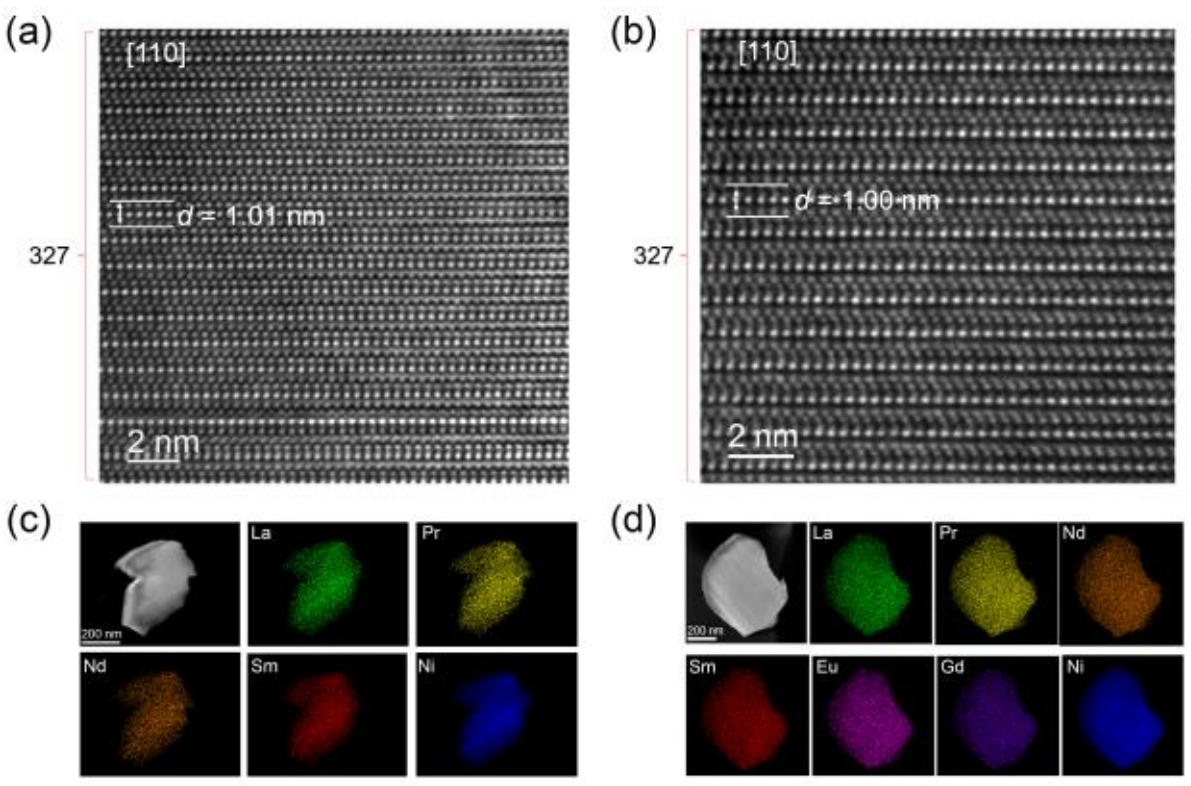

**Figure 2.** (a, b) HRTEM images of ME- and HE-327 along the [110] zone axis, and measured *d*-spacing of (002) planes is about 1.01 and 1.00 nm, respectively. (c, d) HAADF images with corresponding EDS mappings of ME- and HE-327.

The pronounced contraction in the *c* axis reflects significant structural distortions. The 327 structure can be viewed as the alternating stacking of rocksalt *A*O and perovskite $A_2Ni_2O_6$ layers along the *c*-axis [inset of Fig. 1(a)]. The thickness of *A*O layers decreases by 4.9%, from 2.851(4) Å in La-327 to 2.713(5) Å in HE-327. Meanwhile, the thickness of $A_2Ni_2O_6$ bilayers decreases by 1.4%, from 7.411(7) Å in La-327 to 7.307(7) Å. The result suggests smaller $A^{3+}$ ions preferentially occupy the *A*2 site. Within the $A_2Ni_2O_6$ bilayers, the interlayer Ni–Ni distance ($d_\perp$) along the *c* direction is remarkedly reduced by 2.2%, from 4.03(2) Å to 3.94(2) Å. The result suggests enhancement of the coupling between the interlayer Ni-$d_{z^2}$ orbitals, which could be relevant to the DW ordering at ambient pressure and superconductivity under high pressure.

Figs 2(a, b) present HRTEM images of the ME- and HE-327 samples. The sharp lattice fringes and absence of stacking faults vividly indicate good crystallinity, which is align with the XRD results. The lattice parameters *c*, derived from *d*-spacings of (002) planes, are in agreement with those from XRD. Complementary HAADF images and corresponding EDS mappings [Figs 2(c, d)], together with SEM-EDX maps [Fig. S2 in SI], demonstrate homogeneous distribution of the rare-earth elements. This confirms the formation of ME- and HE-327 phases at a microscopic level.

The oxygen content is crucial in governing the physical properties. To determine the oxygen stoichiometry, we performed TGA measurements on ME- and HE-327 under reduction atmosphere. The TGA results, shown in Fig. S3, directly yield the oxygen contents of 6.98(1) for ME-327 and 6.97(2) for HE-327. These values are in agreement with the oxygen occupancies of 6.95(5) and 6.94(9) derived from the XRD refinement, providing compelling evidence for the oxygen stoichiometry in both samples.

Fig. 3(a) displays electrical resistivity measurements of the polycrystalline specimen at ambient pressure. Compared with La-327, the ME- and HE-327 samples exhibit worse conductivity with semiconducting-like behavior in the entire measured temperature range. Similar reports can be seen in other rare-earth-doped La-327 with insignificant oxygen vacancy [12,14]. The semiconducting-like behavior observed here suggests a common localization of charge carriers, possibly associated with multiple intertwined localization mechanisms. The small values of $\bar{r}_A$ enhance lattice distortions, leading to bandwidth narrowing. Additionally, the high configurational entropy at the *A*-site introduces local structural randomness and chemical disorder that may act as scattering centers, reducing carrier mobility.

Notably, there is an anomaly in the $\rho(T)$ curves. For unalloyed La-327, the anomaly occurs at $T_{DW} \approx 144$ K, as identified in its derivative curve [Fig. 3(b)]. This feature has been widely reported and discussed in terms of DW transition [1,14,18,32-36]. In particular, spin-density-wave (SDW) transition near 150 K was concluded by nuclear magnetic resonance [34] and muon spin rotation/relaxation [35,36] studies on La-327. Using the same criterion for identifying the DW transition [18], we found that the DW transition temperature ($T_{\mathrm{DW}}$) drastically increases to 159 and 168 K for ME- and HE-327, respectively. Note that the increase in $T_{\mathrm{DW}}$ with chemical pressure is consistent with the physical pressure effect on $T_{\mathrm{DW}}$ [34,36].

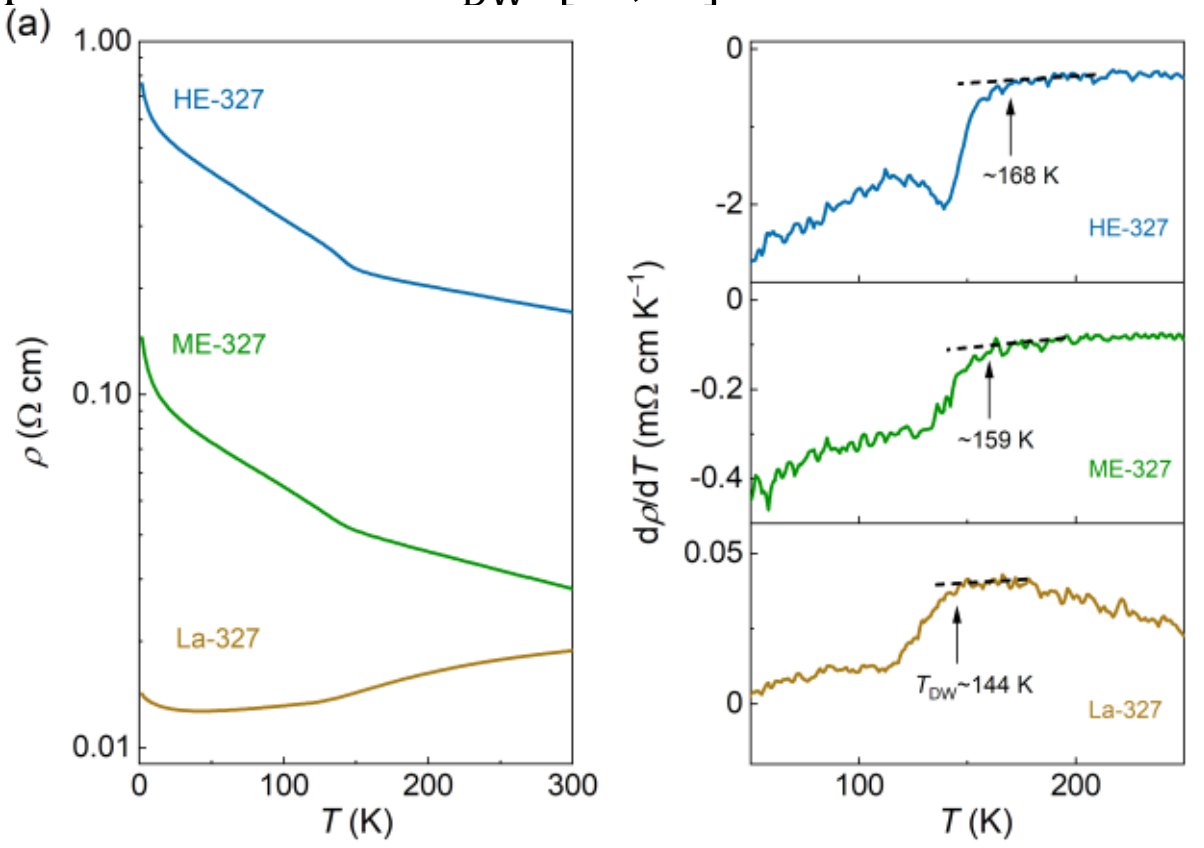

**Figure 3.** (a) Temperature dependence of electrical resistivity of La-, ME-, and HE-327 polycrystalline samples. (b) The corresponding derivatives of resistivity, from which the density-wave transition temperature is identified (indicated by arrows).

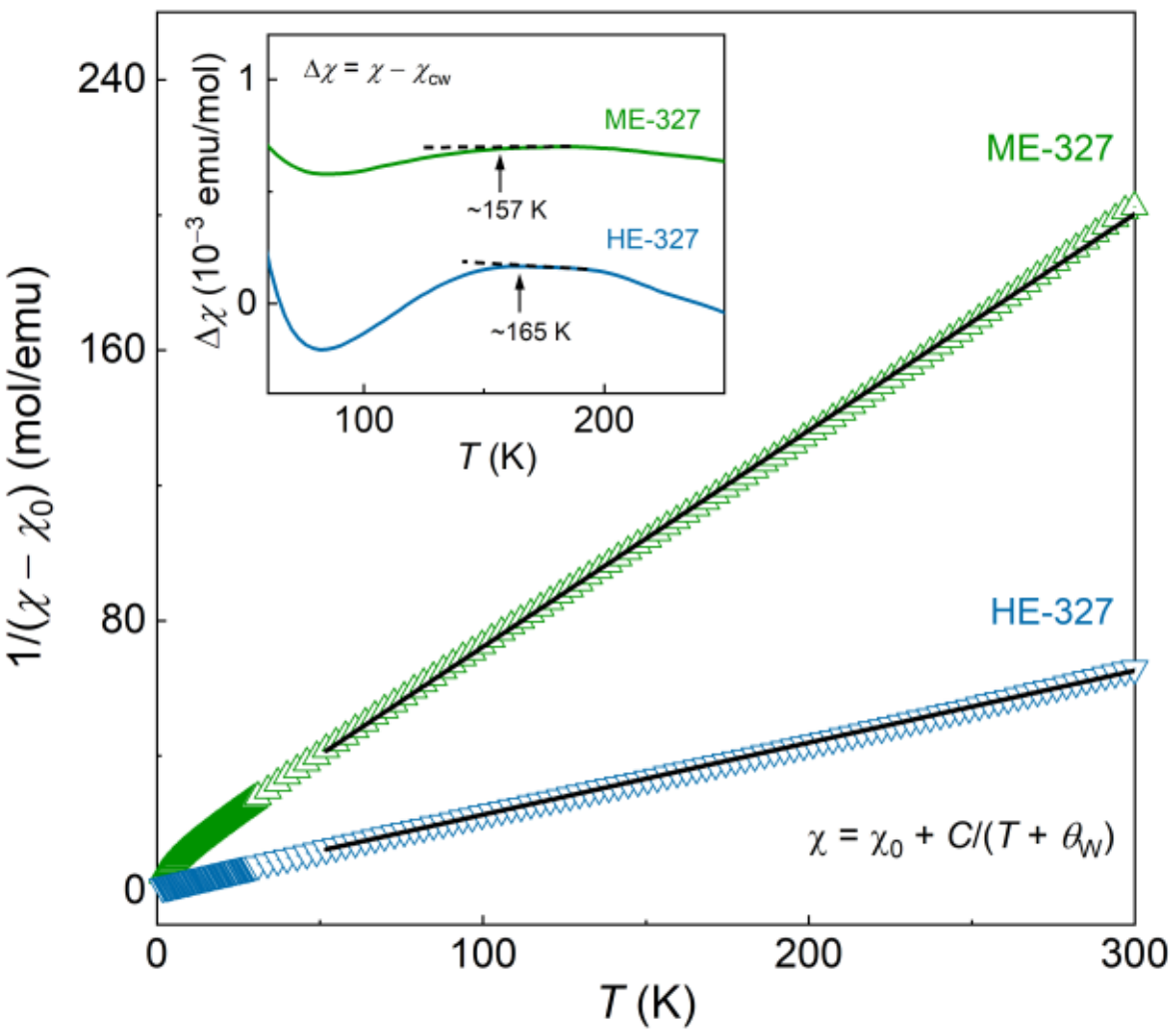

**Figure 4.** Magnetic susceptibility data for La-, ME-, and HE-327, plotted in their reciprocals, $1/(\chi - \chi_0)$. The solid lines denote the Curie-Weiss fit. The inset displays the susceptibility subtracted by the Curie-Weiss fitting, $\Delta\chi = \chi - \chi_{\mathrm{CW}}$. An offset is made for ME-327.

Fig. 4 shows the temperature dependence of the reciprocal of magnetic susceptibility for the ME- and HE-327 samples. Different from La-327 [13,32], these $\chi(T)$ data follow the extended Curie-Weiss formula, $\chi = \chi_0 + C/(T + \theta_W)$, where $\chi_0$, $C$, and $\theta_W$ represent the temperature-independent term, the Curie constant, and the paramagnetic Curie-Weiss (CW) temperature. Obviously, the CW paramagnetism comes from the local moments of magnetic rare-earth ions. The CW fitting (the fitted parameters are presented in Table S3) yields effective magnetic moments of 3.49 $\mu_B$/f.u. (for ME-327) and 6.14 $\mu_B$/f.u. (for HE-327), which reasonably agrees with the theoretically expected values (3.99 $\mu_B$/f.u. for ME-327 and 6.2 $\mu_B$/f.u. for HE-327). To extract the magnetic contributions from Ni ions, we subtracted the CW-fitted curves from the experimental data. The result shows a broad hump at 150-200 K (inset of Fig. 4). The residual susceptibility starts to decrease at ~157 and ~165 K for ME- and HE-327, respectively, consistent with the resistivity measurement result above.

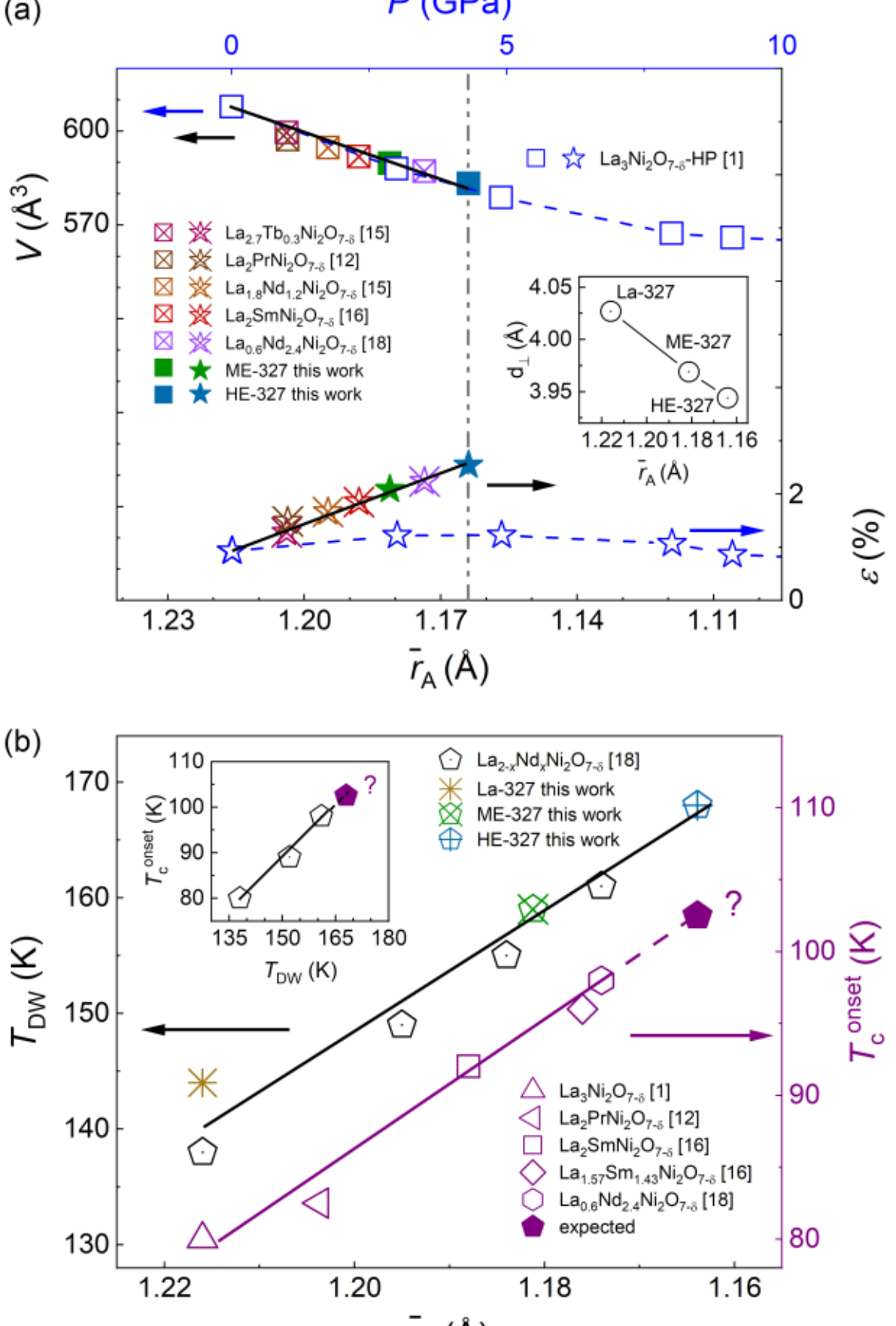

**Figure 5.** (a) Unit cell volume $V$ (left axis) and orthorhombicity $\varepsilon$ (right axis) as functions of $\bar{r}_{\mathrm{A}}$ (bottom axis) and physical pressure $P$ (upper axis), across the series of bilayer nickelates. The inset shows the correlation between $\bar{r}_{\mathrm{A}}$ and the interlayer Ni-Ni distance. (b) DW transition temperature ($T_{\mathrm{DW}}$, left axis) and the onset superconducting transition temperature ($T_c^{\mathrm{onset}}$, right axis) with decreasing $\bar{r}_{\mathrm{A}}$. The prediction $T_c^{\mathrm{onset}}$ of HE-327 is based on an extrapolation of existing high-pressure data from related systems.

Fig. 5(a) plots the unit-cell volume $V$ (left axis) and orthorhombicity $\varepsilon$ [right axis, defined by $\varepsilon = 2(b - a)/(b + a) \times 100\%$] as functions of $\bar{r}_{\mathrm{A}}$ (a measure of chemical pressure) and physical pressure in $A$-327 systems. Various $A$-site doped compounds ($La_{3-x}Ln_xNi_2O_7$, Ln = Tb, Pr, Nd, Sm [12,15,16,18]) and our ME- and HE-327 samples exhibit a systematic decrease in $V$ with decreasing $\bar{r}_A$, in analog with the physical pressure effect in La-327 [1]. By comparing the change in $V$, we estimate that the chemical pressure achieved in HE-327 is equivalent to a physical pressure of ~4.3 GPa. Nevertheless, the chemical pressure exerts significantly different effects on the internal crystal structure. Unlike the physical pressure that hardly changes $\varepsilon$ below 10 GPa and drives the system toward a tetragonal structure at higher pressures, the chemical pressure progressively increases the orthorhombic distortion due to the decrease in the tolerance factor $\tau = \frac{\bar{r}_A + r_O}{\sqrt{2}(r_{Ni} + r_O)}$ [37]. Furthermore, chemical pressure also shortens the interlayer Ni–Ni distance $d_\perp$ as shown in Fig. 5(a). In this circumstance, the in-plane Ni-O-Ni band angles are expected to decrease significantly (here we do not discuss the result quantitatively because the oxygen positions cannot be precisely determined by XRD). Such structural distortion generally leads to band narrowing, which likely leads to carrier localization.

Fig. 5(b) summaries $T_{DW}$ at ambient pressure and the onset superconducting transition temperature $T_c^{onset}$ under high pressures as functions of $\bar{r}_A$. Both $T_{DW}$ and $T_c^{onset}$ increase almost linearly with decreasing $\bar{r}_A$, and our HE-327 sample exhibits the highest $T_{DW}$ value and a high extrapolated $T_c^{onset}$ that exceeds 100 K at high pressure. Our preliminary high-pressure measurements on HE-327 indicated an anomaly in resistivity at 103 K under 31 GPa [Fig. S4], which is a possible signature of superconducting transition. These results align with theoretical results which predicts a higher superconducting transition temperature $T_c$ for reduced $\bar{r}_A$ [20,38]. This is due to the enhancement of interlayer superexchange coupling of $d_{z^2}$ through shortening of $d_\perp$ [39]. Notably, the parallel increase in $T_{DW}$ with $T_c^{onset}$ in inset of Fig. 5(b) suggests a positive correlation between the density-wave order at ambient pressure and the superconductivity under high pressure.

In summary, we have successfully synthesized bilayer nickelates ME-327 and HE-327 with significantly reduced $\bar{r}_A$, thereby bearing a higher chemical pressure. The synthesized polycrystalline samples are phase pure and homogeneous with good crystallinity. Among the bilayer nickelates reported to date, the HE-327 sample possesses the lowest cell volume, the largest orthorhombicity, and shortened interlayer Ni-Ni distance. The physical property measurements indicate that HE-327 exhibits low electrical conductivity and high density-wave transition temperature. Shortening in $d_\perp$ probably promotes the interlayer coupling, which is widely believed to be crucial for superconducting pairing under pressure. The superconducting transition temperature $T_c$ under high pressure increases continuously with chemical pressure, exceeding 100 K for our HE-327 by extrapolation. Our work demonstrates the ionic size effect, revealing clear correlations between the average ionic radius $\bar{r}_A$, orthorhombicity $\varepsilon$, $T_{DW}$ at ambient pressure, and $T_c$ under high pressure. The HE approach provides a new avenue for developing superconducting bilayer nickelates for the future. For example, growth of HE nickelates in single-crystalline and thin-film forms would be crucial for exploring intrinsic anisotropic properties and potential device applications.

**Acknowledgements** This work was supported by the National Natural Science Foundation of China (12494593, 12494592), the CAS Superconducting Research Project under Grant No. [SCZX-01011] and the National Key Research and Development Program of China (2022YFA1403202 and 2023YFA1406101).

Supplementary Information

# Entropy stabilization and effect of *A*-site ionic size in bilayer nickelates

Jia-Yi Lu,[1] Jia-Xin Li,[1] Xin-Yu Zhao,[1,2] Ya-Nan Zhang[1,2], Yi-Qiang Lin,[1] Kai-Xin Ye[1,2], Hui-Qiu Yuan[1,2,3,4] and Guang-Han Cao[1,3,4]*

[1]School of Physics, Zhejiang University, Hangzhou 310058, China

[2]New Cornerstone Science Laboratory, Center for Correlated Matter and School of Physics, Zhejiang University, Hangzhou, China.

[3]Institute of Fundamental and Transdisciplinary Research, and State Key Laboratory of Silicon and Advanced Semiconductor Materials, Zhejiang University, Hangzhou 310058, China

[4]Collaborative Innovation Centre of Advanced Microstructures, Nanjing University, Nanjing, 210093, China

*Correspondence to: ghcao@zju.edu.cn

## Experimental Details

All polycrystalline samples were synthesized by solid-state reactions using a sol-gel produced precursor. Stoichiometric mixture of the source materials, $A(NO_3)_3\cdot 6H_2O$ ($A$ = La, Pr, Nd, Sm, Eu, Gd; 99.9% Aladdin), $Ni(NO_3)_2\cdot 6H_2O$ (99.99% Aladdin), were dissolved in the glycol and deionized water with addition of appropriate amount of citric acid. The mixed solution was continuously stirred on a heating plate (140 °C) for 4 h, and homogeneous green gel resulted. The gel was slowly heated to 500 °C in air, and then further heated to 800 °C, holding for 10 h to eliminate organic components. The resulted precursor was ground and pressed into pellets, and the pellets were sintered at 1100 °C for 50 h in oxygen atmosphere (0.2 MPa). The oxygen gas was generated by the decomposition of $Ag_2O$ (99.9% Aladdin) (together with the sample pellets, an appropriate amount of $Ag_2O$ was placed and sealed in an evacuated silica ampule).

Powder XRD data were collected on a PANalytical diffractometer (Empyrean Series 2) with Cu-$K_{\alpha 1}$ radiation. The crystal structure was refined by Rietveld analysis using the GSAS-II package [1]. The Scanning electron microscopy-energy-dispersive X-ray analysis (SEM-EDX) was performed on a scanning electron microscope (Hitachi S-3700N) equipped with Oxford Instruments X-Max spectrometer. High-resolution transmission electron microscopy (HRTEM) and high-angle annular dark-field (HAADF) imaging with corresponding energy dispersive spectrometer (EDS) mappings were conducted using a FEI Tecnai G² F20 S-Twin scanning transmission electron microscope. Thermogravimetric analysis (TGA) measurements were carried out in a HQT-3 thermal analyzer. The atmosphere was a 10% $H_2$/Ar flow (3 mL/min) and the heating rate was 20 °C /min. Electrical resistivity was measured on a Quantum Design Physical Property Measurement System (PPMS-9). The magnetic properties were measured on a Quantum Design Magnetic Property Measurement System (MPMS3).

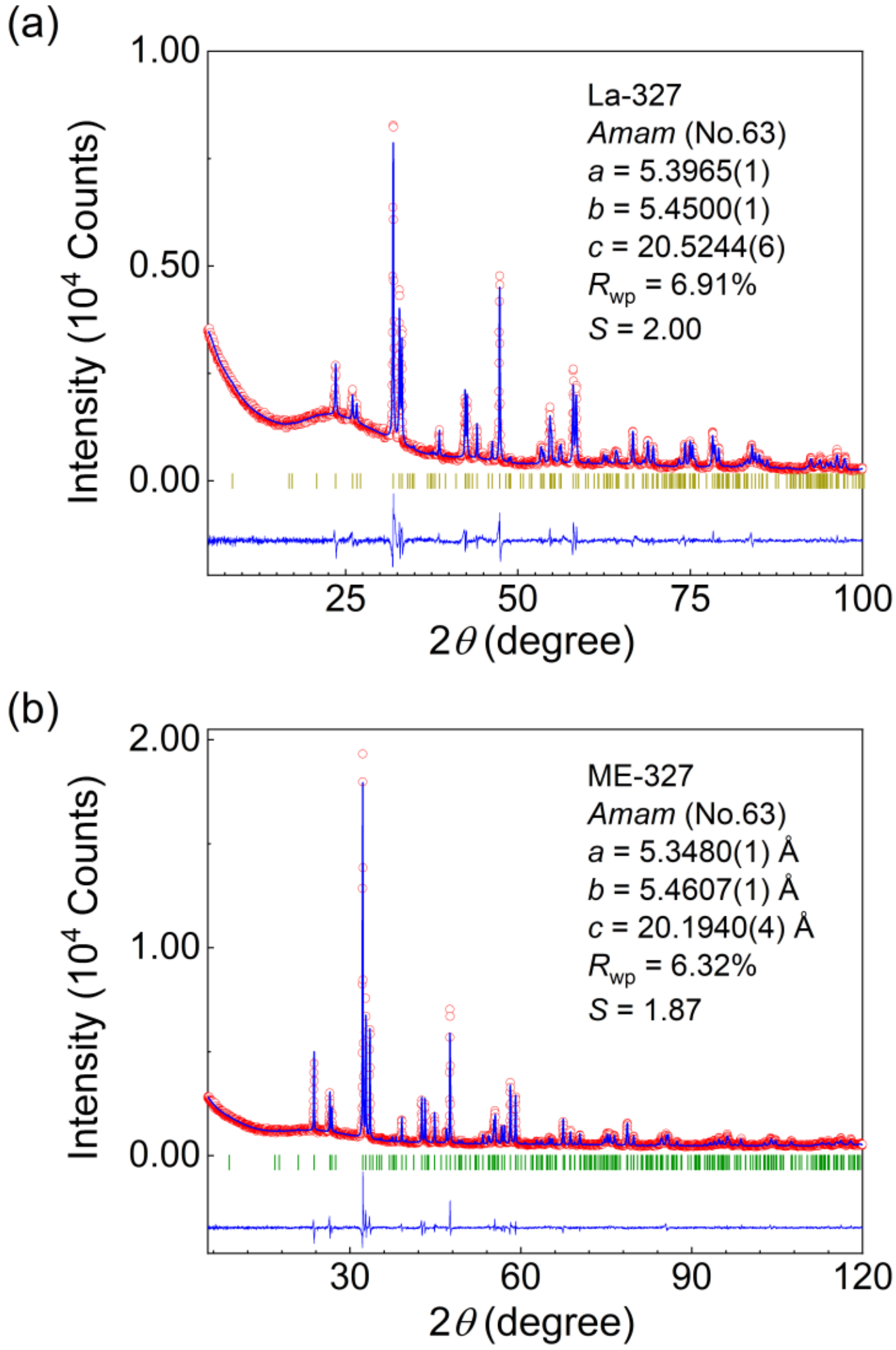

**Figure S1.** Rietveld refinement XRD profiles for La-327 (a) and ME-327 (b). The obtained lattice parameters are $a$ = 5.3965(1) Å, $b$ = 5.4500(1) Å, $c$ = 20.5244(6) Å for La-327 and $a$ = 5.3480(1) Å, $b$ = 5.4607(1) Å, $c$ = 20.1940(4) Å for ME-327.

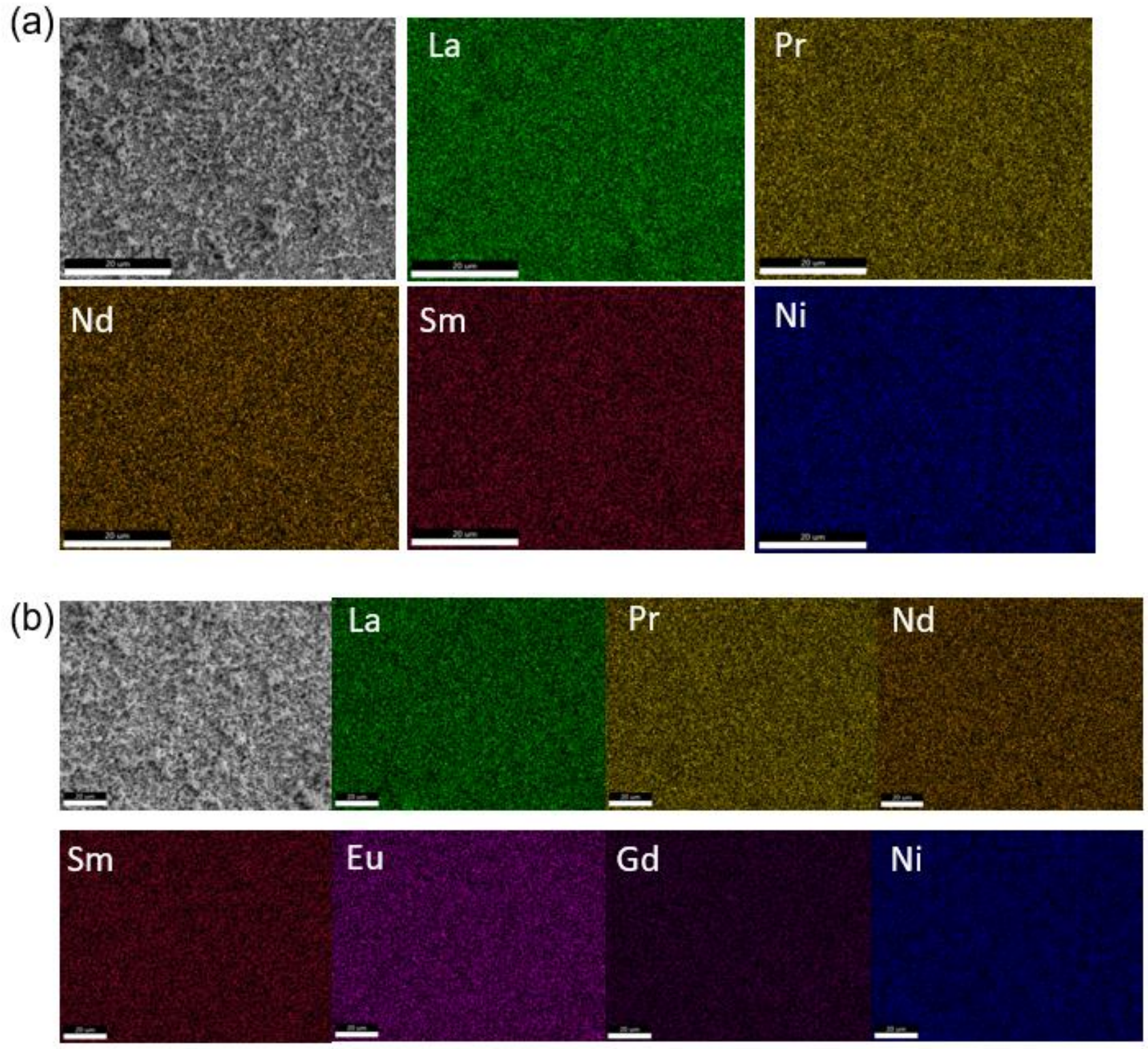

**Figure S2.** SEM-EDX morphology for the ME- and HE-327 samples.

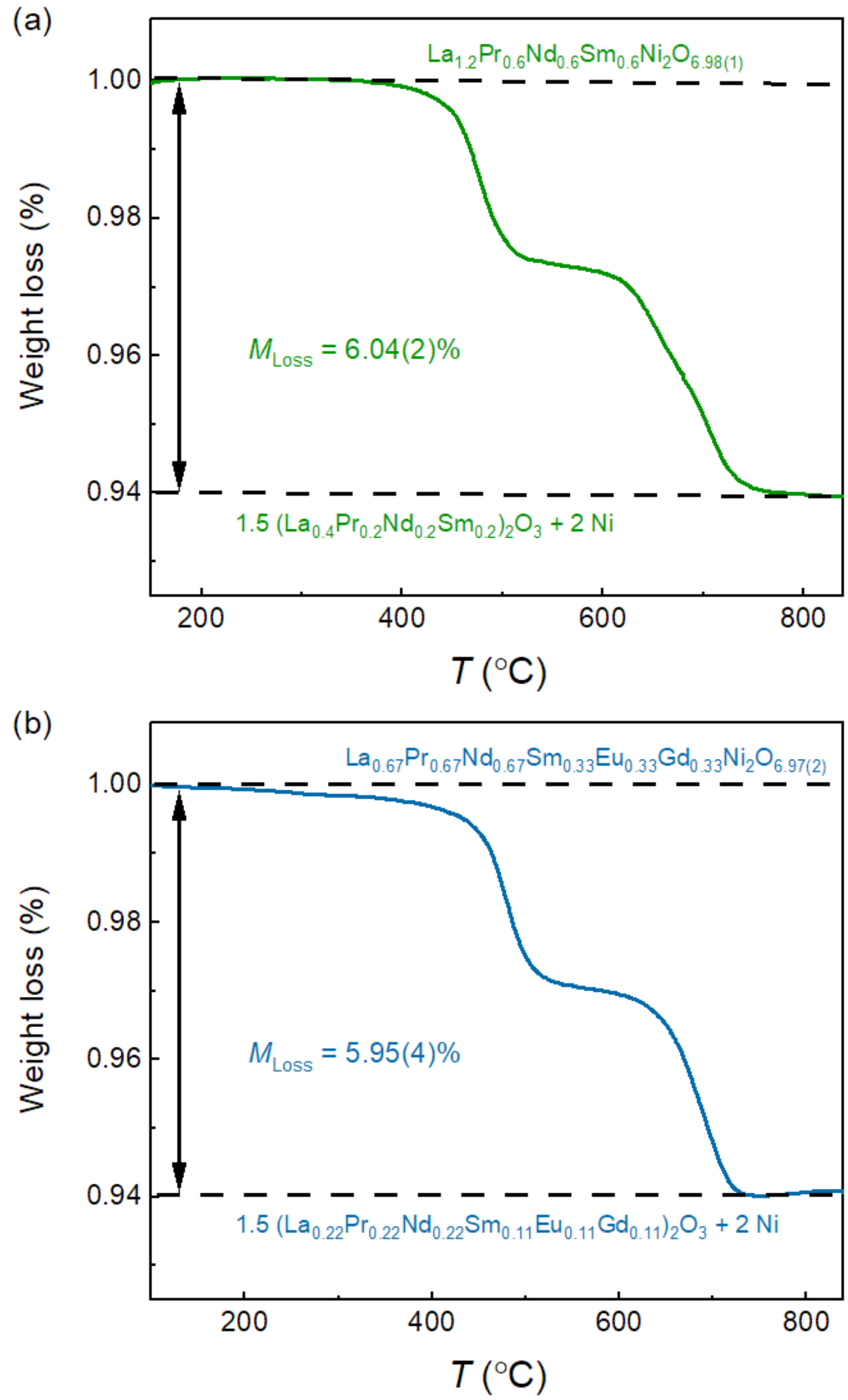

**Figure S3.** Thermogravimetric curves for ME-327 (a) and HE-327 (b) nickelates.

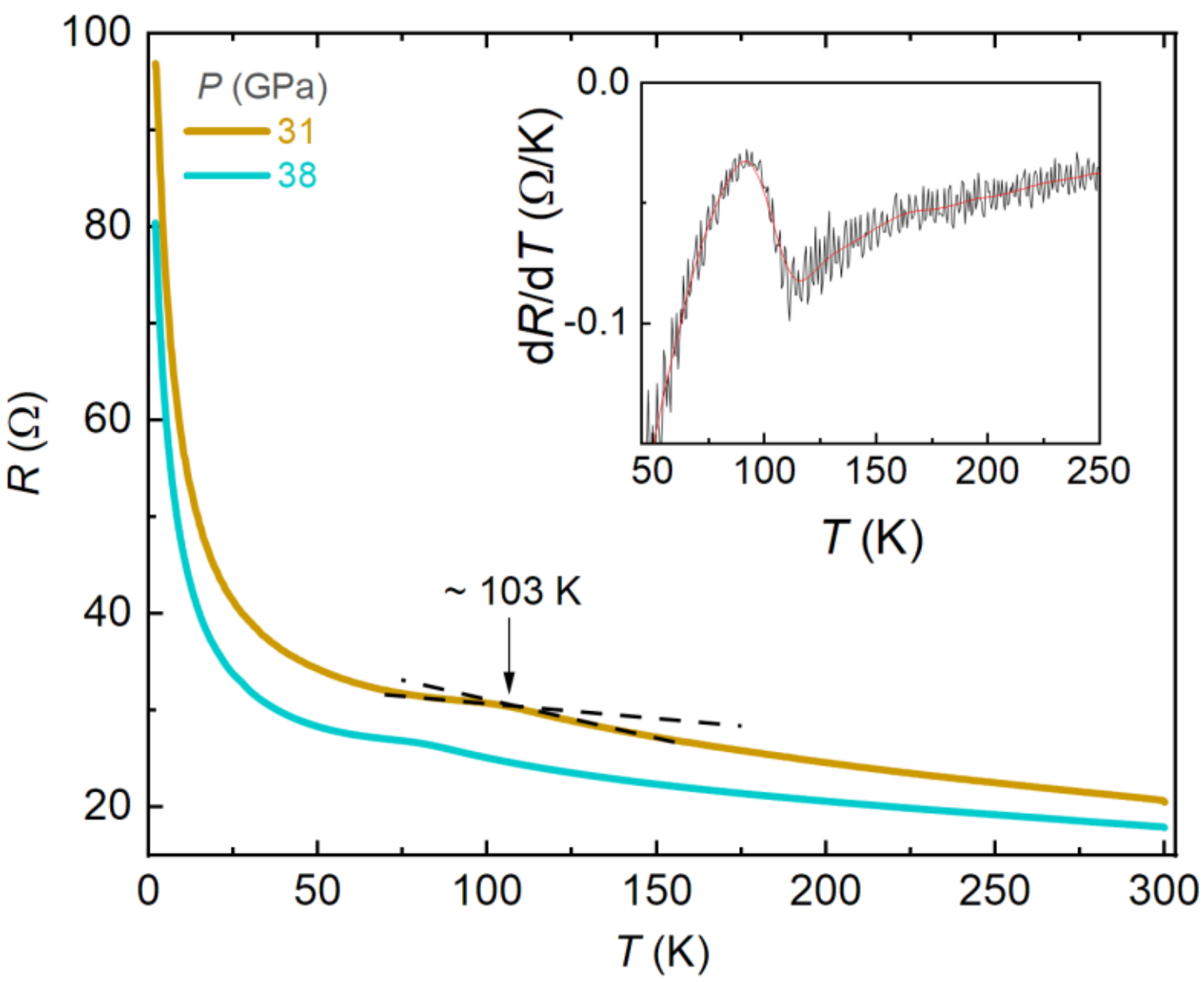

**Figure S4.** Temperature dependence of resistance under pressures for HE-327 sample.

**Table S1.** Configuration entropy for the nickelates of La-, ME-, and HE-327.

| Model | Description | La-327 ($\bar{r}_A$ = 1.216 Å) | ME-327 ($\bar{r}_A$ = 1.181 Å) | HE-327 ($\bar{r}_A$ = 1.164 Å) |
| --- | --- | --- | --- | --- |
| Random | Fully random distribution over both *A* sites. | 0 | 1.33*R* | 1.73*R* |
| Preferential Occupation# | Larger ions preferentially occupy the *A*1 site, according to Ref. [2]. | 0 | 1.21*R* | 1.57*R* |

#: The *A*1 site is preferentially occupied by larger (La, Pr, and Nd) ions randomly to maximize the Gibbs free energy. Therefore, the plausible occupations in $(A1)(A2)_2Ni_2O_7$ are: *A*1 = $(La_{0.33}Pr_{0.33}Nd_{0.33})$ and *A*2 = $(La_{0.44}Pr_{0.14}Nd_{0.14}Sm_{0.3})$ for ME-327, and *A*1 = $(La_{0.33}Pr_{0.33}Nd_{0.33})$ and *A*2 = $(La_{0.17}Pr_{0.17}Nd_{0.17}Sm_{0.17}Eu_{0.17}Gd_{0.17})$ for HE-327.

**Table S2.** Crystallographic data of ME- and HE-327 from XRD Rietveld analysis.

| **ME-327 *Amam* (No. 63)** | |
|---|---|
| *a* (Å) | 5.3480(1) |
| *b* (Å) | 5.4607(1) |
| *c* (Å) | 20.1940(4) |
| *V* (Å$^3$) | 589.74(1) |

| **atom** | ***x*** | ***y*** | ***z*** | **occ.** | $U_{iso}$ |
|---|---|---|---|---|---|
| ***A*1** | 0.25 | 0.250(2) | 0.5 | 1.0 | 0.0063 |
| ***A*2** | 0.25 | 0.263(1) | 0.3182(1) | 1.0 | 0.0063 |
| **Ni** | 0.25 | 0.261(3) | 0.0983(3) | 1.0 | 0.0023 |
| **O1** | 0.25 | 0.295(10) | 0 | 0.95(5) | 0.007 |
| **O2** | 0.25 | 0.184(5) | 0.206(1) | 1.0 | 0.007 |
| **O3** | 0 | 0.5 | 0.112(1) | 1.0 | 0.007 |
| **O4** | 0.5 | 0 | 0.088(2) | 1.0 | 0.007 |

| **HE-327 *Amam* (No. 63)** | |
|---|---|
| *a* (Å) | 5.3266(1) |
| *b* (Å) | 5.4639(1) |
| *c* (Å) | 20.0400(4) |
| *V* (Å$^3$) | 583.24(1) |

| **atom** | ***x*** | ***y*** | ***z*** | **occ.** | $U_{iso}$ |
|---|---|---|---|---|---|
| ***A*1** | 0.25 | 0.259(2) | 0.5 | 1.0 | 0.0063 |
| ***A*2** | 0.25 | 0.261(3) | 0.3177(2) | 1.0 | 0.0063 |
| **Ni** | 0.25 | 0.254(3) | 0.0984(4) | 1.0 | 0.0023 |
| **O1** | 0.25 | 0.364(12) | 0 | 0.94(9) | 0.007 |
| **O2** | 0.25 | 0.156(8) | 0.192(2) | 1.0 | 0.007 |
| **O3** | 0 | 0.5 | 0.108(2) | 1.0 | 0.007 |
| **O4** | 0.5 | 0 | 0.090(2) | 1.0 | 0.007 |

**Table S3.** Parameters of Curie-Weiss fit for ME- and HE-327 nickelates.

| Samples | ME-327 | HE-327 |
|---|---|---|
| $\chi_0$ ($\mathbf{emu \cdot mol^{-1}}$) | 0.0036 | 0.0045 |
| $\theta_W$ (K) | 14.27 | 7.53 |
| $C$ ($\mathbf{emu\ K\ mol^{-1} fu^{-1}}$) | 1.52 | 4.71 |
| $\mu_{eff}$ ($\mu_B$/f.u.) | 3.49 | 6.14 |
| $\mu_A$ ($\mu_B$/f.u.) | 3.99 | 6.20 |

#: The theoretically expected effective magnetic moments are calculated as $\mu_A = \sqrt{\sum_i (x_i \cdot \mu_i^2)}$, where $x_i$ is the molar fraction of each magnetic ion and $\mu_i$ is its free-ion effective moment derived from $\mu_i = g_J\sqrt{J(J+1)}$. Here $g_J$ is the Landé factor and $J$ the total angular momentum of the trivalent ion's Hund's rule ground state. The calculated free-ion values are: $Pr^{3+}$ 3.58 $\mu_B$, $Nd^{3+}$ 3.62 $\mu_B$, $Sm^{3+}$ 0.85 $\mu_B$, $Gd^{3+}$ 7.94 $\mu_B$, with $La^{3+}$ and $Eu^{3+}$ being non-magnetic ($J = 0$).